\documentclass[fleqn,usenatbib]{mnras}

\usepackage{newtxtext,newtxmath}

\usepackage[T1]{fontenc}

\DeclareRobustCommand{\VAN}[3]{#2}
\let\VANthebibliography\thebibliography
\def\thebibliography{\DeclareRobustCommand{\VAN}[3]{##3}\VANthebibliography}

\usepackage{graphicx}	
\usepackage{amsmath}	
\usepackage{tabularx}

\graphicspath{{images}}

\title[The white dwarf cooling sequence in M37]{Exploring the origin of the extended main sequence turn off in M37 through the white dwarf cooling sequence}

\author[M. Griggio et al.]{
M. Griggio,$^{1,2}$\thanks{E-mail: massimo.griggio@inaf.it}
M. Salaris,$^{3,4}$
D. Nardiello,$^{2,5}$
L. R. Bedin, $^{2}$
S. Cassisi,$^{4,6}$
and J. Anderson$^{7}$
\\
$^{1}$Dipartimento di Fisica, Universit\`a di Ferrara, Via Giuseppe Saragat 1, Ferrara I-44122, Italy\\
$^{2}$INAF - Osservatorio Astronomico di Padova, Vicolo dell'Osservatorio 5, Padova I-35122, Italy\\
$^{3}$Astrophysics Research Institute, Liverpool John Moores University, 146 Brownlow Hill, Liverpool L3 5RF, UK\\
$^{4}$INAF - Osservatorio Astronomico di Abruzzo, Via M. Maggini, I-64100 Teramo, Italy\\
$^{5}$Aix Marseille Univ, CNRS, CNES, LAM, Marseille, France\\
$^{6}$INFN - Sezione di Pisa, Largo Pontecorvo 3, 56127 Pisa, Italy\\
$^{7}$Space Telescope Science Institute, 3700 San Martin Drive, Baltimore, MD 21218, USA
}

\date{Accepted 2023 June 15. Received 2023 May 09; in original form 2023 March 22}

\pubyear{2023}

\begin{document}
\label{firstpage}
\pagerange{\pageref{firstpage}--\pageref{lastpage}}
\maketitle

\begin{abstract}
We use new observations from the Canada-France-Hawaii
Telescope to study the white dwarf cooling sequence of the open
cluster M37, a cluster that displays an extended main sequence turn-off and, 
according to a recent photometric analysis, also 
a spread of initial chemical composition. By taking advantage of a first epoch collected in 1999 with the 
same telescope, we have been able to calculate proper motions for sources as faint as $g$\,$\sim$\,26
(about $\sim$\,6 magnitudes fainter than the {\it Gaia} limit), allowing
us to separate cluster members from field stars. 
This has enabled us to isolate a sample of the
white dwarf population of M37, reaching the end of the cooling sequence (at $g$\,$\sim$\,23.5). 
The here-derived atlas and calibrated catalogue of the sources in the field of view is publicly released as supplementary on-line material. 
Finally, we present an exhaustive comparison of the white dwarf luminosity function with theoretical models, which has allowed us to exclude the age-spread scenario as
the main responsible for the extended turnoff seen in the cluster colour-magnitude-diagram.
\end{abstract}

\begin{keywords}
astrometry --
Hertzsprung–Russell and colour–magnitude
diagrams -- open clusters and associations: individual: M37 (NGC\,2099) -- techniques:
photometric -- white dwarfs
\end{keywords}



\section{Introduction}

During the last few years the unprecedented quality of the photometric 
and astrometric data obtained with the {\it Gaia} spacecraft has greatly 
refined our knowledge of the Milky Way open clusters (OCs). 
The OC census has improved through the rejection of thousands of 
misidentified OCs in the literature and the discovery of several hundreds new confirmed OCs \citep[see, e.g.,][for some examples]{cg18, cg}; moreover,  
the improved determination of stellar memberships and orbital parameters has provided us with a better characterisation of individual clusters. 

In this respect, the analysis of the exquisite, high-precision {\it Gaia} colour-magnitude diagrams (CMDs) of {\sl bona fide} members    
of selected OCs, has recently revealed the presence of extended 
main sequence (MS) turn off (TO) regions and broadened MSs, that cannot be originated 
by field contamination, binaries and differential reddening alone \citep[see, e.g.,][and references therein]{bastian18, marino18, cordoni, griggiom37}. 
These features are  
similar to what is observed in the Magellanic 
where star clusters younger than about 2\,Gyr display extended TO regions \citep[see, e.g.,][and references therein]{mackey, mb, goud, pb16}, and clusters younger than $\sim$\,600-700\,Myr display also split MSs
\citep[see, e.g.,][and references therein]{li17, corr17, marino18b}.

Whilst there is mounting evidence that rotation --as opposed to 
an age range among the cluster population-- is the main culprit to 
explain these features in the CMD of both open OCs and Magellanic Cloud clusters 
\citep[see, e.g.,][and references therein]{bastian18, kamann18, kamann20, kamann23},
our photometric analysis of the $\sim$\,500\,Myr old OC M37 (NGC\,2099) -- with an extended TO and no 
split MS -- has targeted a magnitude range populated by stars with convective envelopes, hence 
predicted to be in any case slow rotators,   
disclosing the presence of a 
sizeable initial chemical abundance spread, which may or may not be somehow related to the 
extended TO \citep{griggiom37}. We made 
use of synthetic stellar population and differential colour-colour diagrams using a combination of
{\it Gaia} and {\it Sloan} photometry to show that 
the observed MS colour spread in the high-precision {\it Gaia} Early Data Release~3 \citep[EDR3][]{2021A&A...649A...1G} CMD can only be reproduced by differential reddening and unresolved binaries plus either a metallicity spread $\Delta \rm [Fe/H]$\,$\sim$\,0.15, or a range of initial helium mass fractions $\Delta Y$\,$\sim$\,0.10.
As discussed in \cite{griggiom37}, the existing spectroscopic (high- and medium resolution) measurements of the cluster stars' metallicity 
provide indications both in favour and against the existence of a [Fe/H] spread (in which case our results would point to a sizeable helium abundance spread), but a high-precision 
differential abundance analysis of a consistent sample of cluster stars is needed to 
address this issue spectroscopically.

It is worth noticing that the existence of chemical abundance spreads in low-mass clusters like OCs \citep[M37 has an estimated mass of just 1\,000--1\,500 $M_{\odot}$, see][]{piskunov} is unexpected and hard to explain, and has important implications not only for 
models of cluster formation and the test of stellar models on CMDs of OCs, 
but also for the technique of chemical tagging 
\citep{tagging}, based on the idea that clustering in chemical space can in principle 
associate individual field stars with their birth clusters, assumed 
chemically homogeneous.
If OCs are commonly born with a sizeable internal [Fe/H] range, the suitability of this technique for field stars in the disk of the Milky Way is challenged.

In this paper, we present a new photometric analysis of M37's white dwarf (WD) cooling sequence (CS), which
improves upon earlier results by \citet{kaliraim37} in several ways.  
The area covered by our observations is over three times larger than \citet{kaliraim37},  
who also used the outer regions of their mosaic to estimate field stars contamination, 
which are however now known to host several members stars \citep{griggio}. 
For our field decontamination we have used 
a safer region much further away from the cluster core, and in addition we exploited their data to obtain proper motions with 
a time baseline of 23 years, which allowed us to
determine a sample of WD members.

Taking advantage of this new data we have performed a theoretical analysis of the observed CS to seek for additional constraints on the 
origin of the cluster extended TO and its chemical abundance spread.
The \textit{present-day} low total mass of M37 seems to preclude the 
presence of multiple generations of stars and hence of an age spread according to the scenario presented by  
\citet{goud}, because the cluster should not be able to retain the ejecta of those first-generation stars that can provide material for further episodes of star formation (asymptotic giant branch stars, supernovae). However,  
the chemical composition spread we detected photometrically seems to suggest otherwise, hence it is important to derive 
independent constraints about the origin of the observed extended TO.
The study of the WD cooling sequence and its consistency --or lack of-- with ages inferred from the TO can provide us with these independent clues.

We also publicly release the catalogue with magnitudes and proper motions of the covered region, containing more than 120\,000 sources.

The outline of the paper is as follows. Section\,\ref{obs} presents our new  observations, the data reduction process, and the artificial 
star tests; Sections\,\ref{cs} and \ref{theory} present the observed WD CS and its theoretical analysis, respectively, and are followed by 
Section\,\ref{conclusions} with the conclusions.

\section{Observations}
\label{obs}
\begin{figure}
    \centering
    \includegraphics[width=\columnwidth]{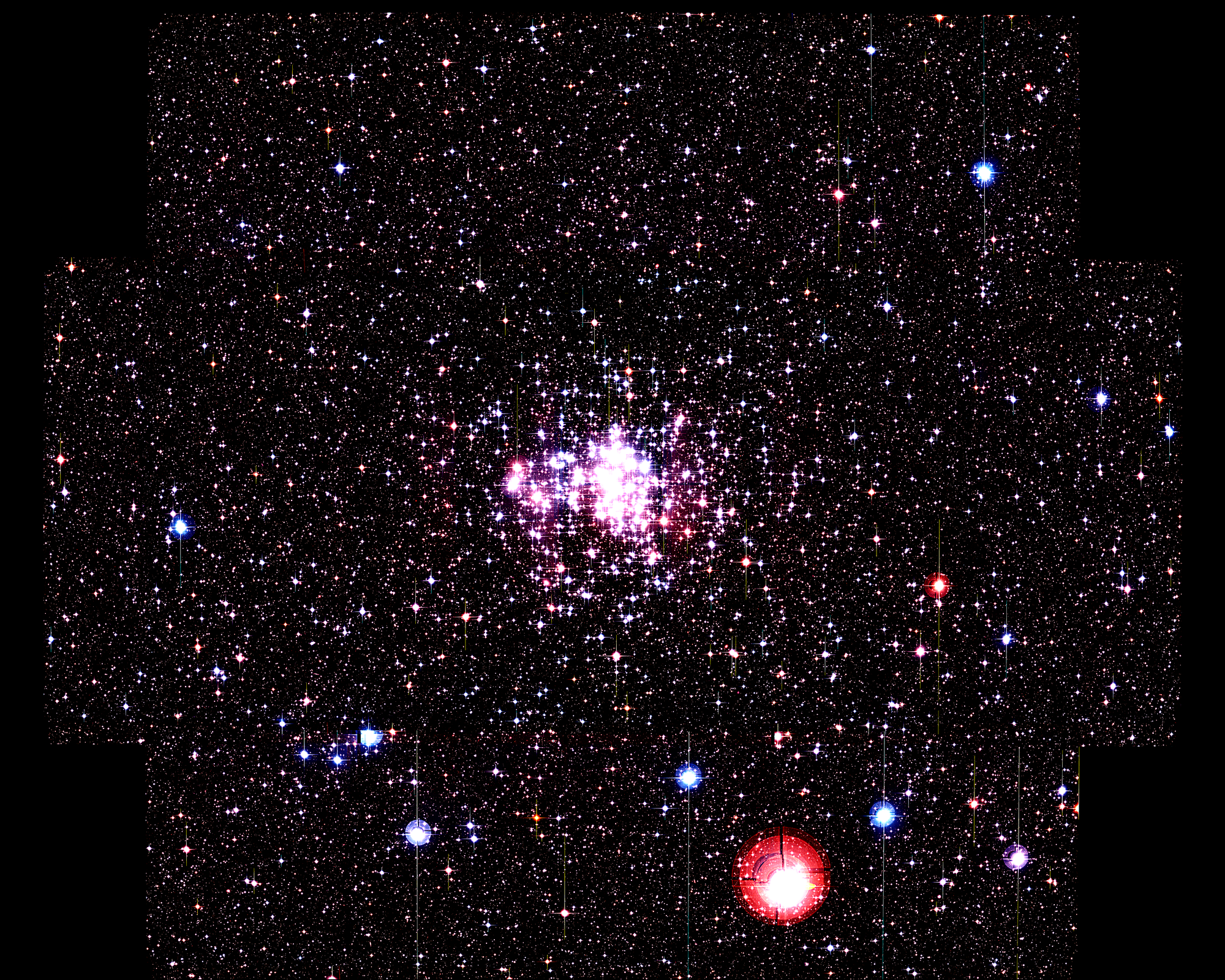}
    \caption{Three-colour view of the field of view. We used $g$ as blue, $r$ as red,
    and a combination of $gr$ for the green colour.}
    \label{fig:stack}
\end{figure}

The main data employed in this article was obtained with the MegaPrime camera at {\it CFHT}, between 
September 27th and 29th, 2022 (PI: Nardiello). The MegaPrime camera is composed of 
forty $2048\times4612$\,pixels CCDs,
with a pixel scale of $\sim$\,0.187\,arcsec/px.
We collected a set of three images with an exposure time of 300\,s, and three 
images of 5\,s, both in the {\it Sloan} filters $g$ and $r$.
The observations in $g$ were repeated twice, for a total of eighteen images, 
twelve in $g$ and six in $r$. 
The data was dithered enough to cover the CCDs' gaps,
with a total field of view of about $1.2\times1.0$\,sq.\,degrees;
a three-colour stacked image of the data is shown in Fig.\,\ref{fig:stack}.

Since the brightest members of M37 MS and all the red clump stars were
saturated even in the short exposures, we collected a set of 50 dithered images with exposure
times of 10\,s in both $g$ and $r$ with the Asiago Schmidt telescope, to complete the
photometry of the brighter part of the CMD. The Asiago Schmidt telescope has 
a $\sim$\,1\,sq.\,degree field of view, and similar data collected with this instrument 
were described in \cite{griggio}.
 
We also took advantage of an early epoch collected at {\it CFHT} (with the pioneering CHF12K camera, 12 CCDs, $\sim$\,0.206\,arcsec/px, $42\times28$\,sq.\,arcmin) in 1999 \citep[PI: Fahlman,][]{2001AJ....122..257K}, to obtain proper motions. 
The CH12K was one of the first wide-field CCD camera to become operative, and these images were collected 
in the {\it Johnson} $B$ and $V$ filters. We used three images per filter, with an exposure time of 300\,s.

A log of the observations is reported in Table \ref{tab:obs}.

\begin{table}
    \caption{Summary of the observations.}
    \label{tab:obs} 
    \centering
    \begin{tabularx}{.9\columnwidth}{lXXX}
        \hline \hline
        Filter & Exp. time & N. of images & Avg. seeing \\
        \hline
        \textbf{Megaprime} & & & \\
        \hline
        $g$ & 300\,s & 6 & 0.55\,arcsec \\
        $g$ &   5\,s & 6 & 0.58\,arcsec \\
        $r$ & 300\,s & 3 & 0.56\,arcsec \\
        $r$ &   5\,s & 3 & 0.70\,arcsec \\
        \hline
        \textbf{Schmidt} & & \\
        \hline
        $g$ & 10\,s & 50 & 1.86\,arcsec\\
        $r$ & 10\,s & 50 & 1.97\,arcsec\\
        \hline
        \textbf{CFH12K} & & \\
        \hline
        $B$ & 300\,s & 3 & 0.79\,arcsec \\
        $V$ & 300\,s & 3 & 0.81\,arcsec \\
        \hline
    \end{tabularx}
\end{table}

\subsection{Preliminary photometry}
As a first step, we derived a \lq{preliminary photometry\rq}, i.e. we
measured the flux and position of the brighter sources, 
that are then used as a starting point to correct for the geometric
distortion and to compute the transformations between the different
exposures.
We treated each CCD of each exposure as an independent image;
in the following we will use the terms \lq{exposure\rq} and \lq{image\rq} 
to refer to the image associated to the single CCD.
Using a version of the
software by \cite{2006A&A...454.1029A} adapted to the {\it CFHT} data,
we computed a $5\times9$ grid of empirical point spread functions (PSFs) for each
image to take into account for the time variations; 
the grid is necessary to account for the spatial variation of the PSF across the CCD. 
Each PSF is derived empirically from bright, unsaturated and isolated
stars, and to each point on the image we associated a local PSF
by a bilinear interpolation of the four closest PSFs in the grid.
We then used the software described in \cite{2006A&A...454.1029A} to
find and measure the position and flux of the sources in the images
by using the local PSF. 
The software outputs a catalogue with positions and instrumental magnitudes of the sources for each exposure.

\subsection{Geometric distortion}

Given that one of our goals was to measure proper motions, we needed accurate positions in both epochs.
To this purpose, we corrected the geometric distortion
following the same approach for both the detectors CFH12K and MegaPrime 
\citep[the procedure is similar to the one adopted in][]{griggio}.

We selected bright ($g_{\rm instr}<-10$), unsaturated sources from each catalogue derived by the preliminary photometry.
We cross-identified the sources in our catalogues with the sources in the {\it Gaia} DR3 catalogues,
projected onto the tangent plane of each image in its central pixel,
after transforming the positions to the epoch of each observation.
We then fitted the residuals between the {\it Gaia} positions and the positions measured in our images 
with a third-order polynomial,
and applied the 75\% of the correction. 
We then repeated the process, starting with the corrected 
positions of the previous iteration, reaching convergence after 30 iterations.

After the correction, the residuals' dispersion for bright sources is smaller than 0.05 pixels in both
detectors, corresponding to $\sim$\,10\,mas for the 1999 data and to $\sim$\,9\,mas for the 2022 data;
summing up these residuals in quadrature we obtain a positional dispersion of $\sim$\,14\,mas, 
to be diluted over a time-baseline of $\sim$\,23\,years, i.e. about 0.6\,mas\,yr$^{-1}$. 
Given the absolute proper motion of M37,
which is about 6\,mas\,yr$^{-1}$ \citep{gb22}, this will allow for a 
proper-motion-based separation between field objects and cluster members 
(see Sec.\,\ref{sec:pho_ast}).

\subsection{Master frame and zero-points calibration}
%
To measure the faintest sources in the field of view, we needed to perform deep
photometry as in \cite{griggio} (which we name \lq{second-pass photometry\rq}, 
see Sec.\,\ref{sec:pho_ast}).
This requires to define
a common reference system for all the exposures, to which we then refer the positions in both epochs,
that we call \lq{master frame\rq}.
The master frame was defined by the positions of the {\it Gaia} DR3 catalogue, projected onto the plane
tangent to the central point of image \texttt{506225p} for CFH12K data,
and \texttt{2785599p} for MegaPrime data. 
The {\it Gaia} positions were again transformed to the epoch of each observation.
We used the catalogues of each image to derive the six-parameter transformations to bring the positions 
measured in the detector reference frame of each exposure onto the corresponding master frame.

The MegaPrime exposures were also dithered enough to allow us measuring the CCDs'
relative photometric zero points, which
we found to be of the order of 0.01\,mag. 
Our derived $BV$ photometry for the CFH12K dataset, however, was not usable, 
in part because we could not access the calibration files, and in part because of the 
non-ideal dither pattern, which did not allow us to register the CCD zero points to 
a common photometric reference system. 
Therefore, the 1999 CFH12K images were used only to derive positions in this first epoch, 
which were in turn employed to derive the proper motions 
necessary to decontaminate cluster stars from field objects.

\subsection{Photometry and astrometry}

To extract the positions and fluxes for all the sources in the field of view we used the code
\texttt{KS2}, an evolution of the code developed by \cite{2008AJ....135.2055A} for 
the Hubble Space Telescope data,
which was adapted to deal with the {\it CFHT} data.

\begin{figure}
    \centering
    \includegraphics[width=\columnwidth]{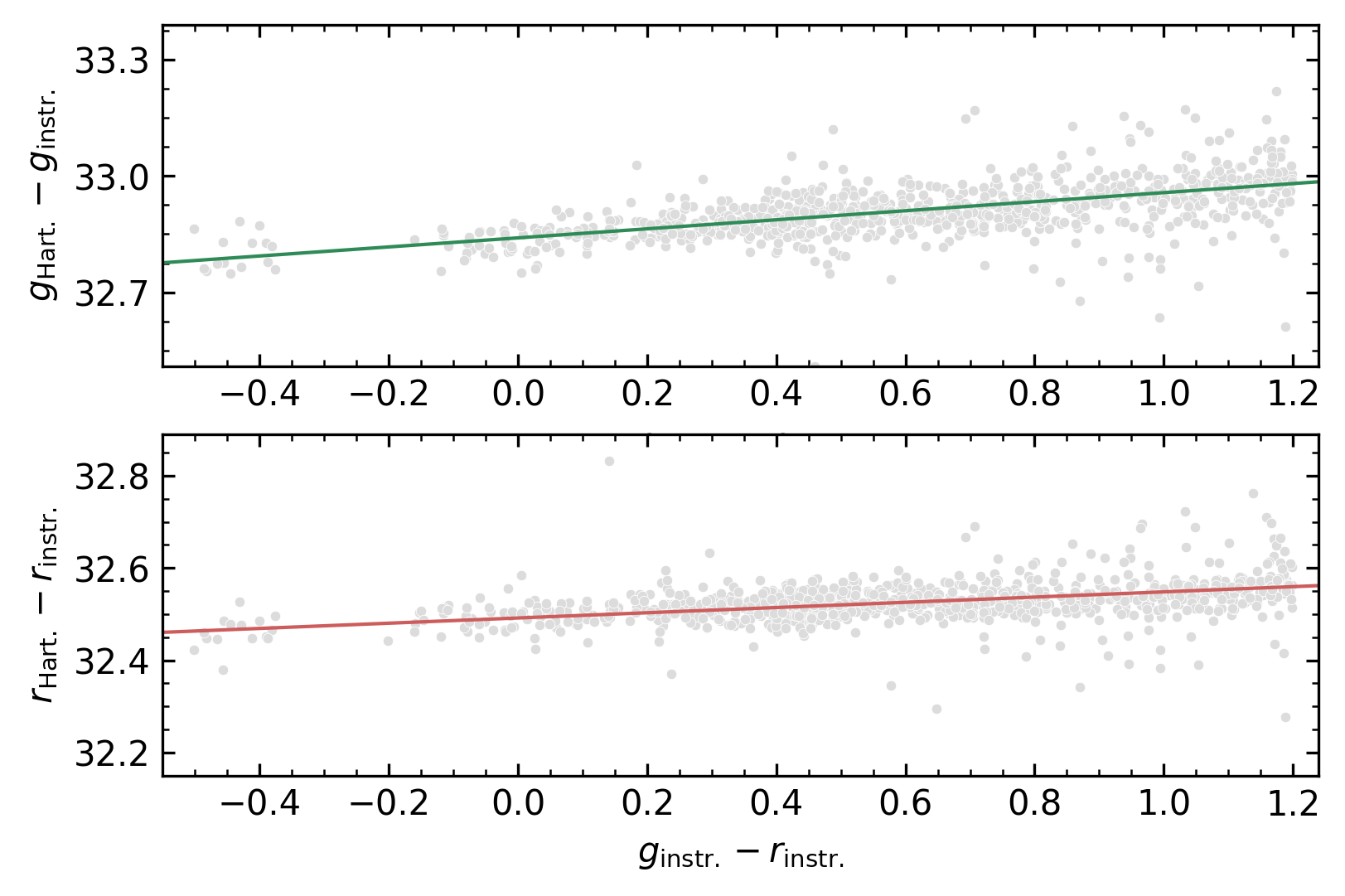}
    \caption{
     Calibration of the {\it CFHT} $gr$ filters: the coloured lines denote the linear fit to the data. 
    We display the difference 
    in the $g$ and $r$ filters between \protect\cite{2008ApJ...675.1233H} and our instrumental magnitudes, as a function of the instrumental $(g-r)$.}
    \label{fig:cal}
\end{figure}

The program goes through several iterations, finding and measuring progressively fainter stars,
using all the images simultaneously to find the sources, thus increasing the signal-to-noise ratio.
This allows to find even the faintest sources that are lost in the noise in single exposure.
The software uses a list of bright stars (derived from the preliminary photometry)
to construct weighted masks, that help to avoid PSF-related artefacts.
The flux is measured performing a PSF fitting of the inner 
$5\times5$ pixels of the source, with the appropriate local PSF, and averaged between all
the images, with a local sky computed from the surrounding pixels.
Measured stars are subtracted from the image before proceeding with the next iteration.
The program outputs also some quality flags \citep[see, e.g.,][]{2009ApJ...697..965B},
that we used to discard sources with galaxy-like shape and diffraction spikes.

\label{sec:pho_ast}
\begin{figure}
    \centering
    \includegraphics[width=.9\columnwidth]{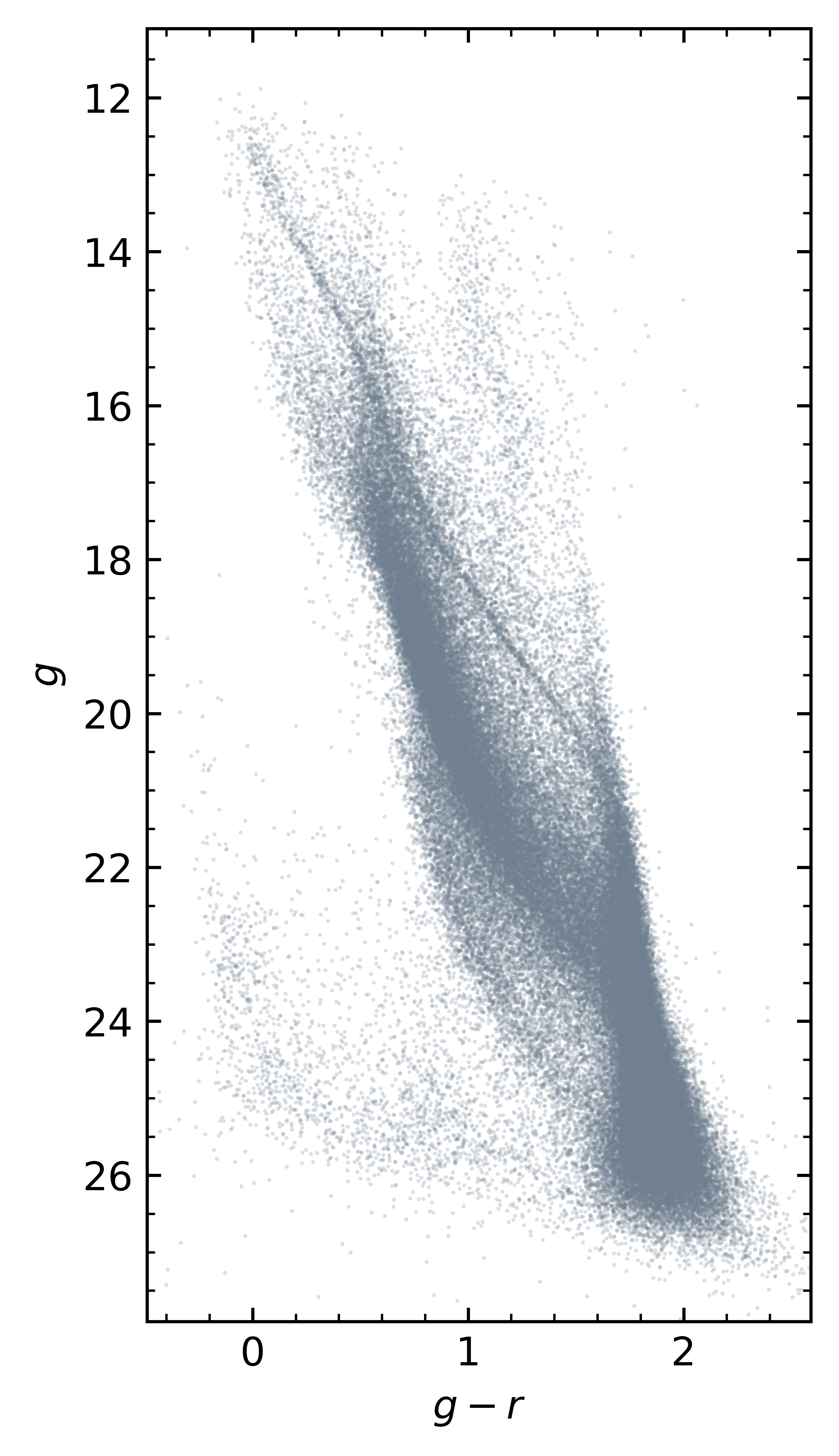}
    \caption{CMD of all the sources that passed the quality cuts in the $gr$ filters ($\sim$\,120\,000).}
    \label{fig:cmd_all}
\end{figure}

The $gr$ instrumental magnitudes have been then calibrated using the deep photometric catalogue by 
\cite{2008ApJ...675.1233H} 
by means of a relation in the form
$m_{\mathrm{cal}}=m_{\mathrm{instr}}+a(g_{\mathrm{instr}}-r_{\mathrm{instr}})+b$, 
with the parameters $a$ and $b$ determined from a linear fit, as shown in Fig.\,\ref{fig:cal}.
The calibrated CMD of all the sources in the field of view is shown in Fig.\,\ref{fig:cmd_all}.

We extracted the photometry from the Asiago data as described 
in Sec.\,3.1 of \cite{griggio}. We did not perform the second-pass 
photometry as we needed only the bright sources. 
We employed the same procedure outlined for the MegaPrime data to calibrate the Asiago photometry.

The flux and position of the sources in the 1999 exposures were extracted with the software \texttt{KS2}.
However, due to the issues described in the previous section, 
we did not carry out the photometric calibration.

\begin{figure}
    \centering
    \includegraphics[width=.9\columnwidth]{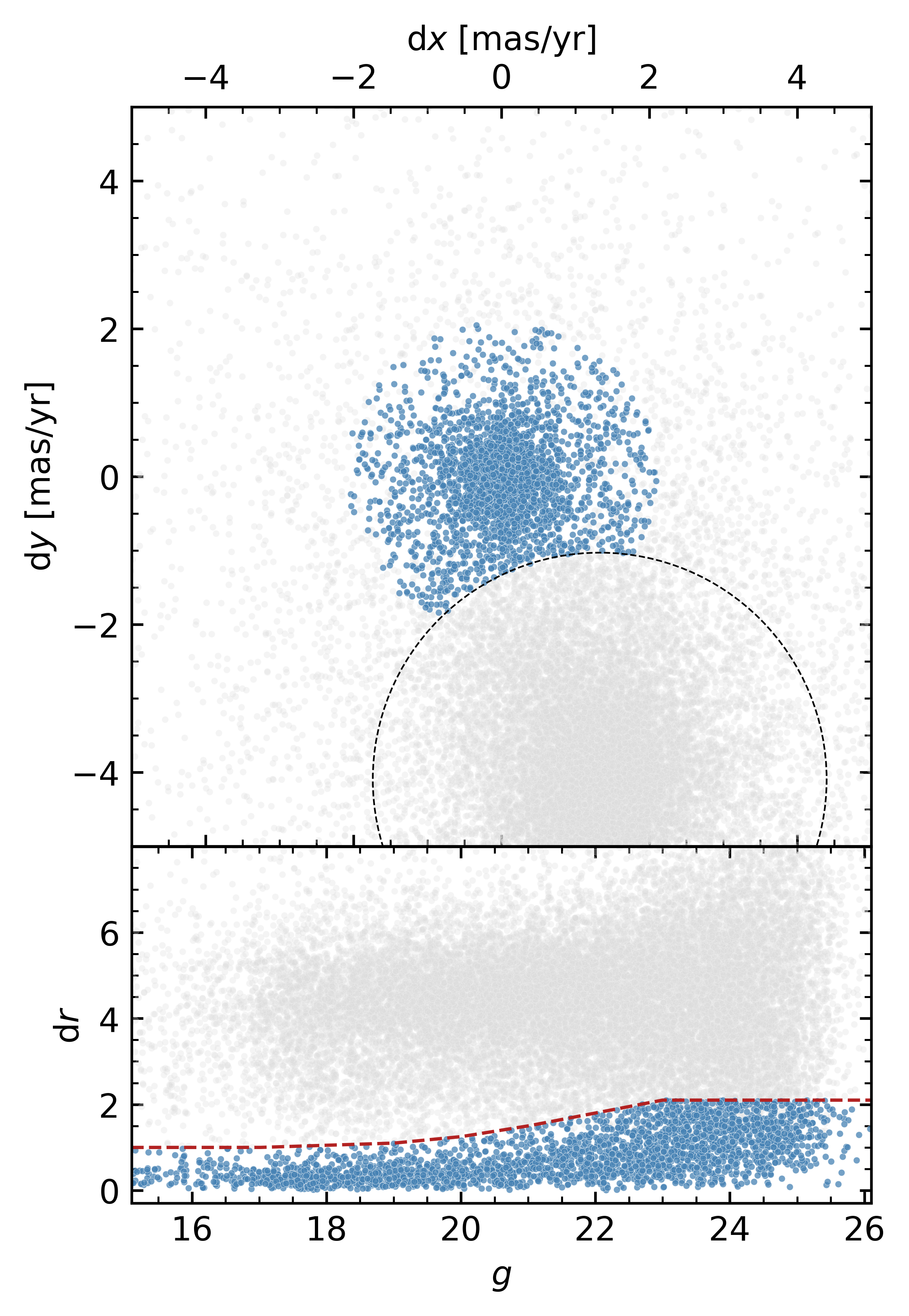}
    \caption{{\sl Top panel}: proper motions for all the sources (grey), with selected members highlighted in blue. The dashed
    black circle is the cut described in the text. The origin is set to the cluster's
    mean proper motion.
    {\sl Bottom panel}: d$r$ vs $g$ for all the sources (grey) and cluster members (blue).}
    \label{fig:pms}
\end{figure}

Proper motions were calculated using the displacements d$x$ and d$y$ between the two epochs, 
divided by the time
baseline of $\sim$\,23\,years, and are shown in Fig.\,\ref{fig:pms} 
(where we used the cluster's mean proper motion as the origin); 
the displacements were measured by transforming
the positions of the stars in the first epoch into the reference system of the second
epoch with a six-parameter transformation, and cross-identifying the common sources. 

The bottom panel of Fig.\,\ref{fig:pms} shows the member selection; we plotted
the distance d$r$ from the origin as function of the $g$ magnitude, and
we drew by hand the red line following the distribution of cluster stars, with a sharp 
cut where cluster and field cannot be well separated by eye. In addition, we estimated the field
median d$x$ and d$y$ and its intrinsic dispersion $\sigma_{x,y}$ as 1.5 times the
68.27$^{\rm th}$ percentile of the distribution of d$x$ and d$y$ around their median, 
and excluded the sources with proper motion inside a circle centred on the field 
motion with radius given by the sum in quadrature of $\sigma_{x,y}$ (dashed black circle
in Fig.\,\ref{fig:pms}).
For sources that are present in the catalogue by \cite{griggio}, we
adopted their \texttt{member} flag, that, for sources at brighter magnitudes, 
is more reliable that the selection based on our 
measured proper motions as it is based on the {\it Gaia} astrometry.

\begin{figure}
    \centering
    \includegraphics[width=.9\columnwidth]{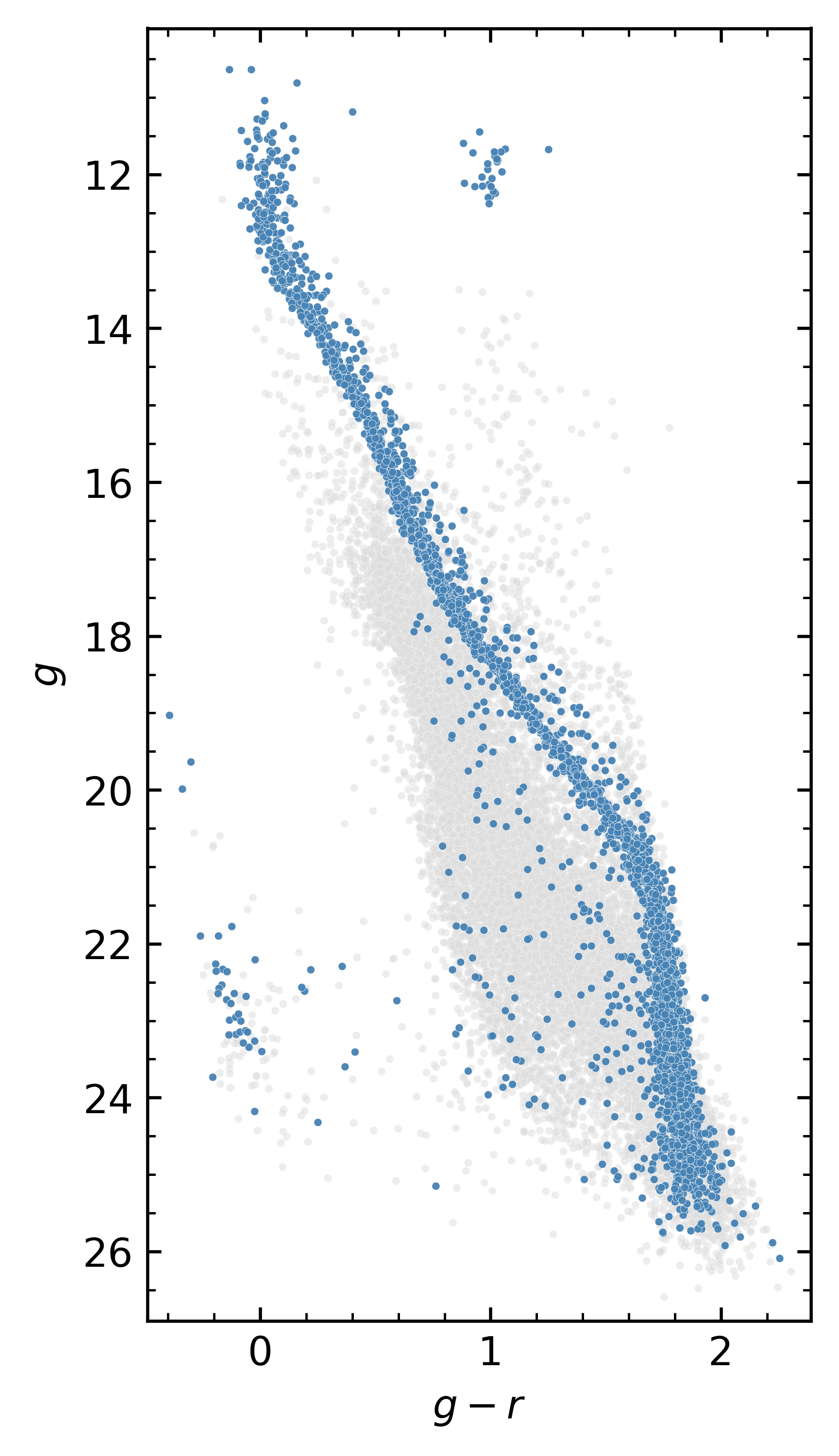}
    \caption{CMD for all the sources with proper motions (light grey, $\sim$\,24\,000) and for those selected as cluster members (blue, $\sim$\,3\,200).}
    \label{fig:cmd_members}
\end{figure}

This selection leads to Fig.\,\ref{fig:cmd_members}, where we show in light grey all the sources
with proper motions (which are less than those in Fig.\,\ref{fig:cmd_all}, as the 2022 data are
deeper and cover a larger area than the 1999 ones) and in blue the selected cluster members. We plotted the {\it CFHT} photometry
up to $g=12.5$, and the Schmidt data for $g<12.5$ to complete the TO and red clump regions which are saturated in the {\it CFHT} short exposures.

Our derived proper motions represent an extension of the {\it Gaia} astrometry down to
$g$\,$\sim$\,26, and the deepest astro-photometric catalogue of M37 available until now. 
Unfortunately, given
the large errors on the positions of faint sources in the first epoch, we cannot discriminate very well between
members and field stars for $g\gtrapprox22.5$.
Nonetheless, we proved the capability of ground-based wide-field
imagers in providing useful astrometry even in the {\it Gaia} era.

Finally, we confirm WD1, WD2 and WD3 of \cite{griggio} as member candidates according to their proper motions obtained in this work,
while WD5 proper motions are not compatible with those of the cluster. The other WDs, namely WD4, WD6 and WD7, fall outside
the field of view, and we could not measure their motion.

\subsection{Artificial stars test}
\label{sec:as}
\begin{figure}
    \centering
    \includegraphics[width=\columnwidth]{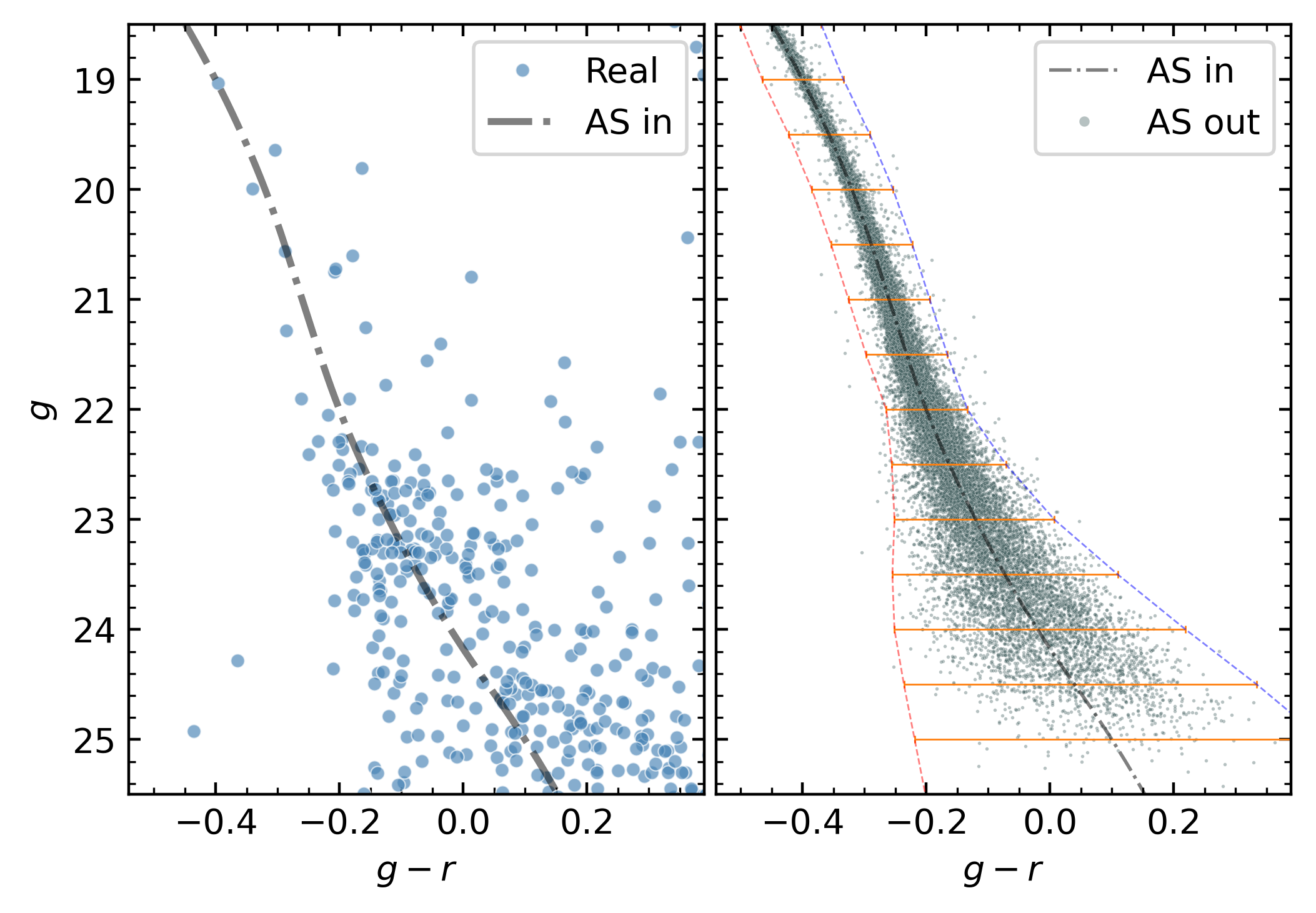}
    \caption{Artificial stars test. \textit{Left panel}: blue points denote the observed white dwarfs, the dark grey line represents
    the fiducial along which we
    generated the artificial stars. \textit{Right panel}: recovered artificial stars. The orange error bars are calculated 
    as three times the 68.27$^{\rm th}$
    percentile of the colour residuals around the median, in each 0.5 magnitude bin. The dashed lines connecting the edges 
    of the error bars define
    the region in which we will count the white dwarfs.}
    \label{fig:as_test}
\end{figure}

\begin{figure}
    \centering
    \includegraphics[width=\columnwidth]{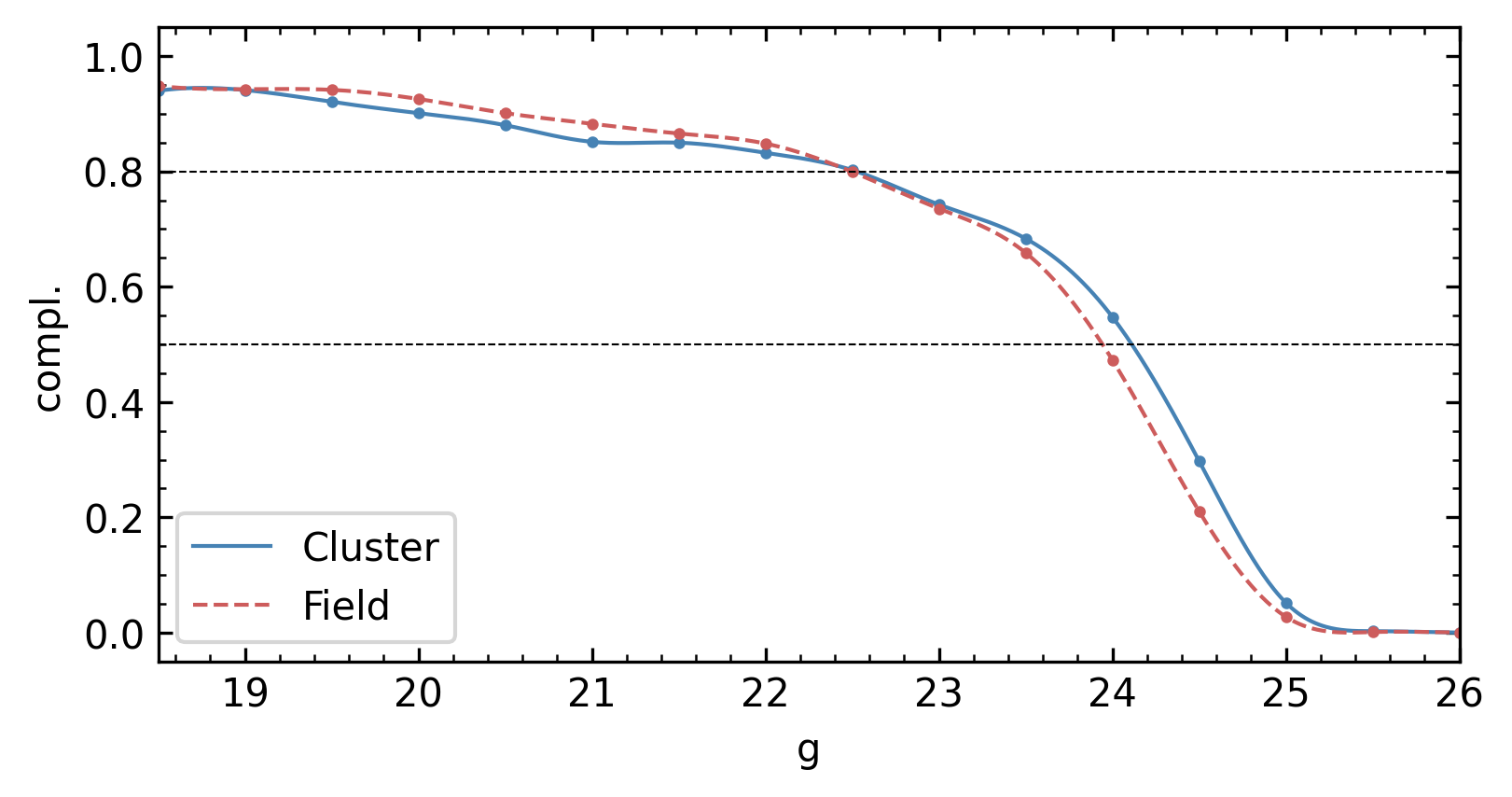}
    \caption{Completeness of our data in the \lq{cluster\rq} and \lq{field\rq} regions.
    See text and Fig.~\ref{fig:fov}.}
    \label{fig:compl}
\end{figure}

To assess the completeness of our data set, we performed the artificial star (AS) test with the
\texttt{KS2} program \citep[see, e.g.,][]{2009ApJ...697..965B}. Briefly, we injected in the images 100\,000 
synthetic stars (one at a time, in order to not create false over-crowding), generated with random positions and random $g$
magnitudes, both sampled from a uniform distribution, with $r$ magnitudes such that they lie on the WD CS fiducial drawn by hand on the CMD 
(Fig.\,\ref{fig:as_test}, left panel).
The software then operates blindly, finding and measuring all the sources in the images. 
We then compared the list of measured stars with the AS input list.
We considered an AS as recovered if its measured position is within 1 pixel in $x$ and $y$ from the injected
position and its magnitude within 0.1 from
the injected magnitudes in both filters.

In the right panel of Fig.\,\ref{fig:as_test} we show the CMD of the recovered stars, 
that guided the choice of the region we adopted to derive the 
WD differential luminosity function (LF).
We divided into 0.5 $g$-magnitude bins the recovered ASs, and for each bin we 
computed the median colour and the $\sigma=68.27^{\rm th}$ percentile of the colour residuals around the
median. The orange error bars in the right panel
of Fig.\,\ref{fig:as_test} represents the $3\sigma$ interval, and the blue and red 
curves connecting the edges of the error bars define the region that we will use for our analysis.

The AS test let us infer the completeness of our data set, defined as the ratio between
the number of recovered stars and the number
of injected stars, which varies across the magnitude range covered by our observations. 
We computed this ratio for each 0.25 $g$-magnitude interval,
and interpolated the values with a spline. The derived completeness curves are plotted in Fig.\,\ref{fig:compl}:
the two horizontal lines mark the 80\% and 50\% completeness levels. Notice that the completeness drops below 50\% at about $g$\,$\sim$\,24, and
reaches zero at $g$\,$\sim$\,26.

\begin{figure}
    \centering
    \includegraphics[width=\columnwidth]{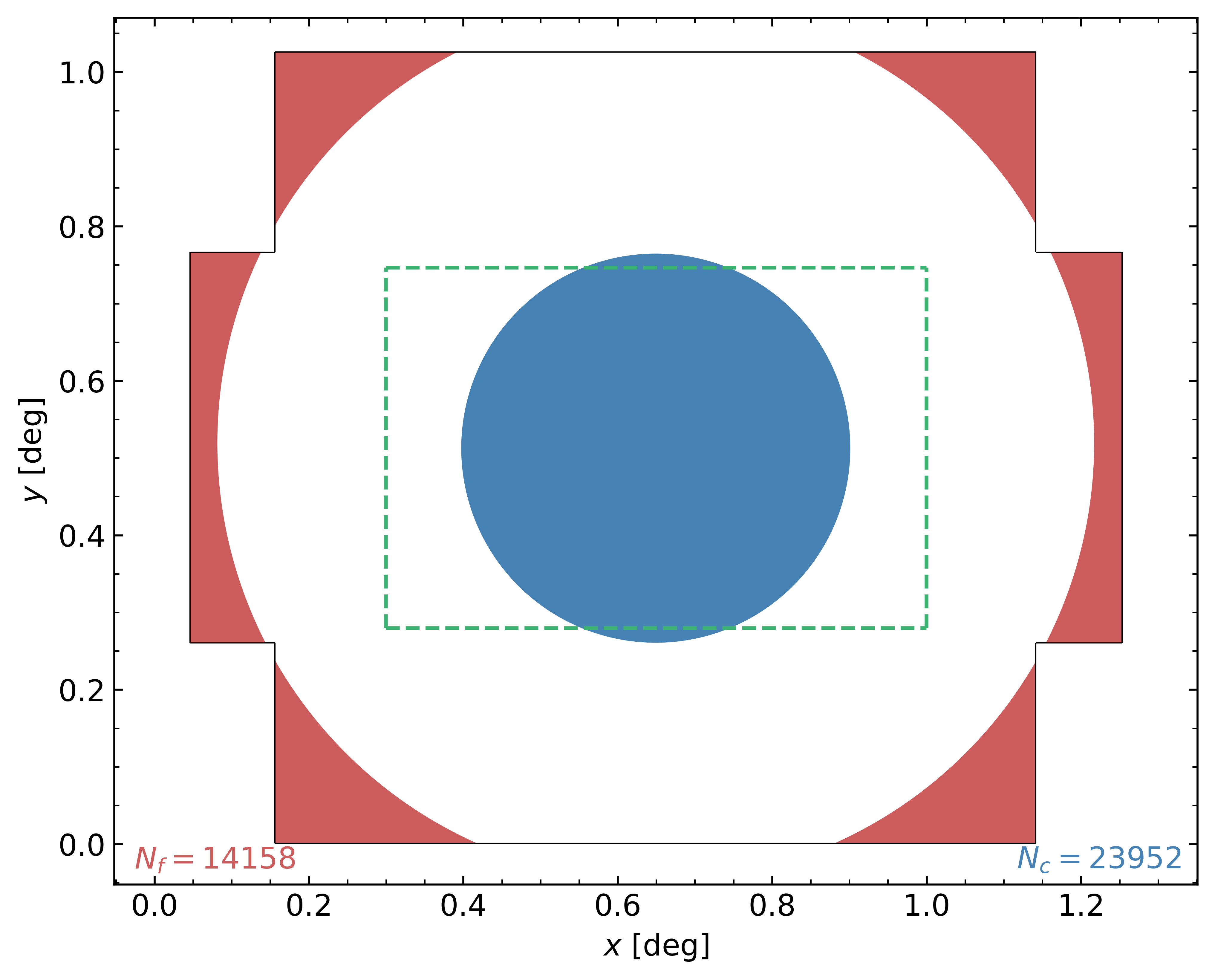}
    \caption{Total field of view of the MegaPrime data. 
    The black lines delimit the observed region. 
    The blue filled area is the ``cluster'' region, while the red filled area is 
    the ``field'' region.
    The two regions have the same area of $\sim$\,70\,Mpx. The number of stars in each region
    is annotated in the lower corners.
    The green dashed line shows the area covered in 1999 by the CFH12K detector.}
    \label{fig:fov}
\end{figure}

The completeness has been computed both for 
the \lq{cluster\rq} region and \lq{field\rq} regions, shown in Fig.\,\ref{fig:fov} in blue and red respectively.
The two regions have roughly the same area of about 0.2\,deg$^2$, and will be employed in Sec.\,\ref{cs} in the study of the WD CS.
 
\subsection{Astro-photometric catalogue}

Together with this work we publicly release an astro-photometric catalogue of the sources that we measured in the {\it CFHT} field of view.
Proper motions are available only for sources in the common region of the 1999 data, which is about one-third of the new dataset 
(as shown in Fig.\,\ref{fig:fov}).

The catalogue contains $x$ and $y$ positions on the master frame in MegaPrime pixels, with $187$\,mas\,px\,$^{-1}$,
the $gr$ photometry and proper motions along
the $x$ and $y$ axes in mas\,yr\,$^{-1}$. In addition, the \texttt{quality} flag denotes sources that passed our quality cuts, 
and the \texttt{member} flag those who are selected as member candidates in this work (blue points in Fig.\,\ref{fig:pms}).

\section{The white dwarf cooling sequence}
\label{cs}

\begin{figure}
    \centering
    \includegraphics[width=\columnwidth]{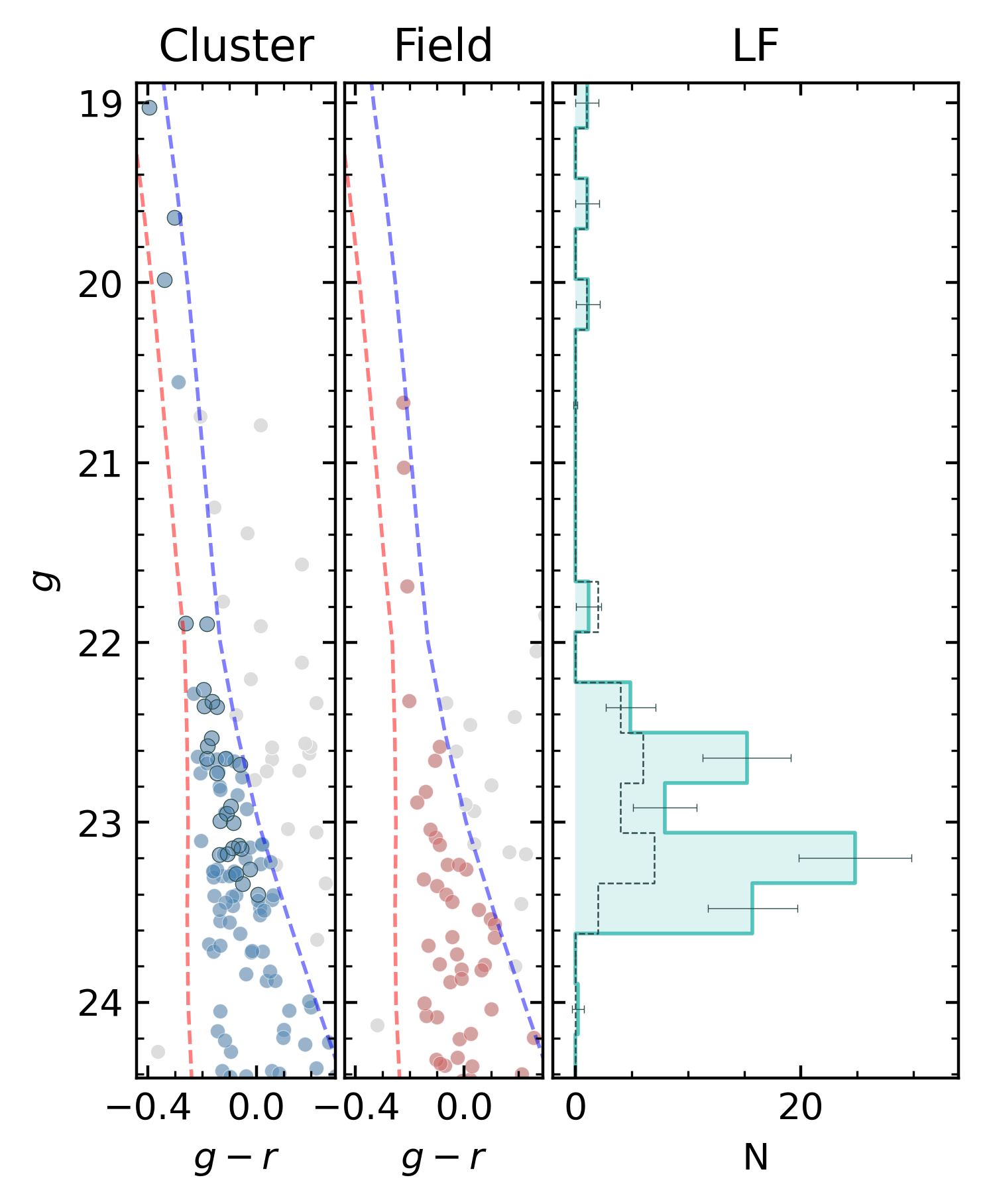}
    \caption{CMD of the WD CS in the \lq{cluster\rq} ({\sl left panel}) and
    \lq{field\rq} ({\sl middle panel}) regions. Blue points with solid black
    edge in the left panel denote the sources that are member candidates
    according to their proper motions. The red and blue dashed lines
    are those defined by the AS test (see Fig.\,\ref{fig:as_test}).
    The {\sl right panel} shows  
    the completeness-corrected
    LF after field decontamination (sea green). The dark-grey dashed line represents
    the LF of the proper motion selected WD members.
    See text for details.}
    \label{fig:lf}
\end{figure}

\begin{table}
    \caption{Our derived, completeness-corrected, WD differential LF. Negative values
    have been set to zero.}
    \label{tab:lf}
    \centering
    \begin{tabularx}{.9\columnwidth}{XXX}
    \hline \hline
    $g$ & N & $\sigma_{\rm N}$ \\
    \hline
    19.00 &  1.1 & 1.0 \\
    19.28 &  0.0 & 0.0 \\
    19.56 &  1.1 & 1.0 \\
    19.84 &  0.0 & 0.0 \\
    20.12 &  1.1 & 1.1 \\
    20.40 &  0.0 & 0.0 \\
    20.68 &  0.0 & 0.2 \\
    20.96 &  0.0 & 0.0 \\
    21.24 &  0.0 & 0.0 \\
    21.52 &  0.0 & 0.0 \\
    21.80 &  1.2 & 1.1 \\
    22.08 &  0.0 & 0.0 \\
    22.36 &  4.9 & 2.2 \\
    22.64 & 15.2 & 3.9 \\
    22.92 &  7.9 & 2.8 \\
    23.20 & 24.8 & 5.0 \\
    23.48 & 15.7 & 4.0 \\
    23.76 &  0.0 & 0.0 \\
    24.04 &  0.3 & 0.5 \\
    24.32 &  0.0 & 0.0 \\
    24.60 &  0.0 & 0.0 \\
    24.88 &  0.0 & 0.0 \\
    \hline
    \end{tabularx}
\end{table}

The 1999 dataset allowed us to measure proper motions for sources well beyond the
{\it Gaia} magnitude limit, down to $g$\,$\sim$\,26; however, given the large errors,
we could not discriminate well between
cluster and field stars at magnitudes fainter than $g$\,$\sim$\,22.5 
(see Fig.\,\ref{fig:pms}, bottom panel).
Most faint sources that have a clear point-like shape in the 2022 data
are heavily affected by the noise in the 1999 data, making their position
(and consequently, their proper motion) measurements very uncertain.
For this reason, we did not employ proper motions to remove field objects
in the derivation of the LF;
we have instead performed a statistical decontamination \citep[cfr.][]{2023MNRAS.518.3722B}
using the regions defined in Fig.\,\ref{fig:fov} to obtain the WD LF that we will compare
to theoretical predictions in the next section. 

Fig.\,\ref{fig:lf} shows the CMD of the WD CS, for both the
\lq{cluster\rq} and \lq{field\rq} regions defined in Fig.\,\ref{fig:fov}. 
The red and blue lines in these CMDs 
are those defined by the AS test (Fig.\,\ref{fig:as_test}) and mark the boundaries 
of the region within which we count WD candidates.
The final LF is given by the difference between the completeness-corrected
\lq{cluster\rq} and \lq{field\rq} LFs, and is shown in sea-green in the right 
panel of Fig.\,\ref{fig:lf} (and reported in Table\,\ref{tab:lf}), with
error bars corresponding to Poisson errors. 
The dashed dark-grey line represents the LF of
WD member candidates selected by proper motions: we note that the two LFs have similar features,
and in particular they terminate at the same magnitude $g$\,$\sim$\,23.5, where the completeness 
level is still greater than 50\% (cfr. Fig.\,\ref{fig:compl}).
This cut-off of the LF is well-defined and can be used as an age indicator for the cluster; for an 
increasing age of the cluster's population, the oldest (earlier forming) WDs 
have more time to cool down, thus shifting the LF cut-off towards fainter magnitudes.

\section{Comparison with theory}
\label{theory}

In this section we discuss the comparison of the WD LF of Table\,\ref{tab:lf} with theoretical WD models,
that enabled us to derive important constraints on the origin of the extended TO observed 
in the cluster CMD.
Due to the issue highlighted below, we have only compared the LF in the $g$ band with theory,
and not the CS in the CMD.

As already mentioned, we found in \citet{griggiom37} that the stellar population hosted by this 
cluster displays either a range of metallicity $\Delta \rm [Fe/H]$\,$\sim$\,0.15 and 
a range of differential reddening $\Delta E(B-V)=0.06$ ([Fe/H] spread scenario), or a 
spread of helium abundance $\Delta Y$\,$\sim$\,0.06 and a range $\Delta E(B-V)=0.03$ (helium spread scenario).
For the distance -- 1450\,pc, consistent with the range $1500\pm100$\,pc 
determined by \citet{gb22} from {\it Gaia} EDR3 parallaxes -- and reference [Fe/H]\,$=0.06$ 
-- consistent with the existing few high-resolution spectroscopic measurements \citep{pancino} -- used in \citet{griggiom37} analysis, 
in the [Fe/H] spread scenario the reference $E(B-V)$ ranges from 0.28\,mag to 0.34\,mag and the metallicity ranges from $\rm [Fe/H]=0.06$ to $\rm [Fe/H]=0.21$. The lowest metallicity 
isochrone \citep[from the BaSTI-IAC database,][]{basti} matches the blue envelope of the unevolved MS in the {\it Gaia} CMD ($G$ magnitudes between $\sim$\,15 and $\sim$\,17) for the lowest value of the reddening, $E(B-V)=0.28$ \citep[see, e.g., Figs. 1 and 9 in][]{griggiom37}.
In the $Y$ spread scenario, we found $Y$ ranging from $Y=0.269$ -- the standard value of $Y$ at
$\rm [Fe/H]=0.06$ in the BaSTI-IAC isochrones -- to $Y=0.369$, for $E(B-V)$ between 0.33\,mag and 0.36\,mag.
In this case, the blue envelope of the unevolved MS in the {\it Gaia} CMD is matched by the most
helium-rich $Y=0.369$ (hence bluer) isochrones, and $E(B-V)=0.33$.

\begin{figure*}
    \centering
    \includegraphics[width=\textwidth]{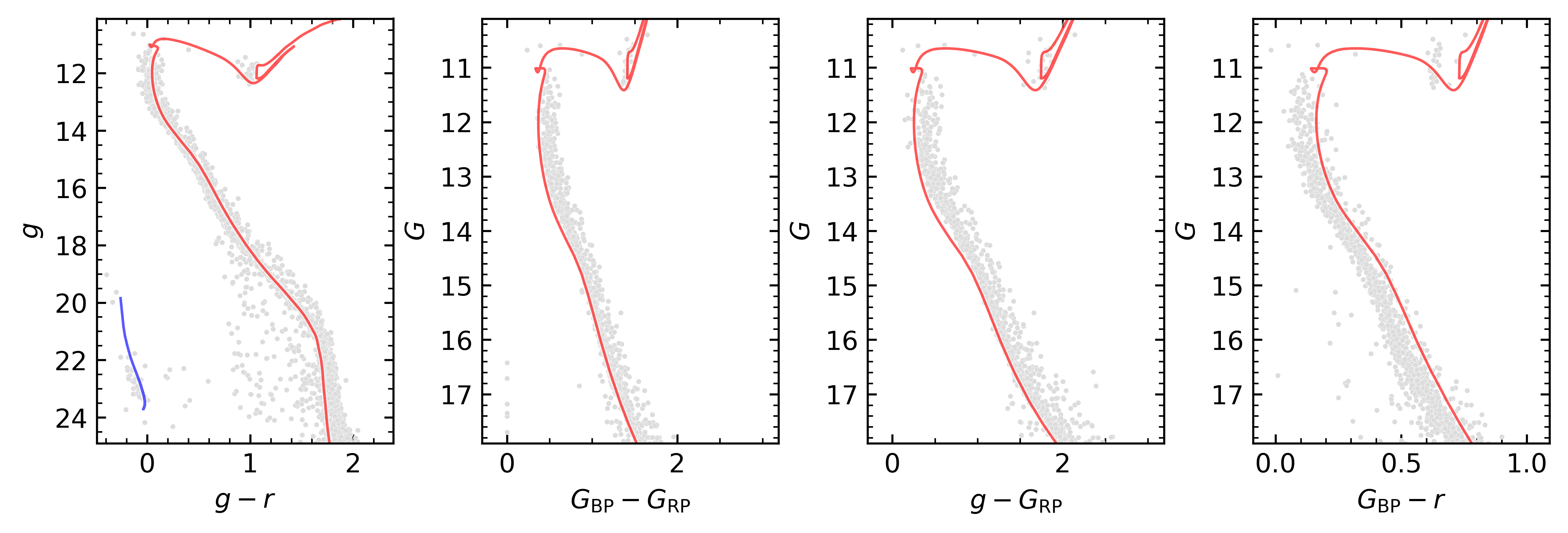}
    \caption{CMDs of M37 stars in several magnitude and colour combinations. Theoretical isochrones (including the WD sequence in the left panel) are compared to the observations using $E(B-V)=0.28$, and a distance $d=1450$\,pc (see text for details). The extinction law is taken from \citet{zhang2023}, for the {\it Sloan} filters and from the {\it Gaia} website (\url{https://www.cosmos.esa.int/web/gaia/edr3-extinction-law}) for {\it Gaia} magnitudes.}
    \label{fig:cmd_iso}
\end{figure*}

The leftmost panel of Fig.\,\ref{fig:cmd_iso} shows that, in the [Fe/H] spread scenario, when we match a
400\,Myr (the exact age is irrelevant to this discussion) $\rm [Fe/H]=0.06$ BaSTI-IAC isochrone \citep[from the same sets adopted in][]{griggiom37} to the MS in the {\it Sloan} $g$-$(g-r)$ CMD using $E(B-V)=0.28$ and the extinction ratios by \citet{zhang2023}, the models are redder than the blue edge of the unevolved MS in a wide magnitude range. This includes the interval between $g$\,$\sim$\,16 and $\sim$\,19, which approximately corresponds to the $G$ magnitude range of the {\it Gaia} CMD where the isochrones match the blue edge of the MS \citep{griggiom37}, as 
shown is the second panel from the left of the same figure.  

We also show a 400\,Myr WD BaSTI-IAC isochrone calculated from hydrogen-envelope (DA) carbon-oxygen (CO) core WD cooling tracks \citep[computed with the][electron conduction opacities]{c07} with $\rm [Fe/H]=0.06$ progenitors by \citet{bastiwd}, the initial-final-mass relation (IFMR) by \citet{ifmr} and progenitor lifetimes from \citet{basti}, compared to the observed CS for the same choice of distance and reddening. The WD isochrone also appears redder than the observations.

To investigate the cause(s) of this inconsistency with the fit to the MS in the {\it Gaia} CMD, the third and fourth panel from the left in Fig.\,\ref{fig:cmd_iso} display CMDs with the {\it Gaia} $G$ magnitude on the vertical axis, and colours calculated using one {\it Gaia} and one {\it Sloan} magnitude. The same isochrone of the left panel is compared to the data in these two CMDs.
We can see that the models in the $G-(g-G_{\rm RP})$ CMD match the 
blue edge of the MS, whilst the isochrones are redder than the observed MS in the $G-(G_{\rm BP}-r)$ CMD. This suggests that the inconsistency between the fits in the two photometric systems arises from a mismatch between the theoretical and observed $r$ magnitudes\footnote{This mismatch exists also in comparison with \citet{2008ApJ...675.1233H} photometry, which has been used to calibrate our magnitudes, as shown in Fig.\,\ref{fig:cal}.}.
For this reason, in our study of the WD cooling sequence, we will consider only the $g$ magnitudes.

\begin{figure}
    \centering
    \includegraphics[width=\columnwidth]{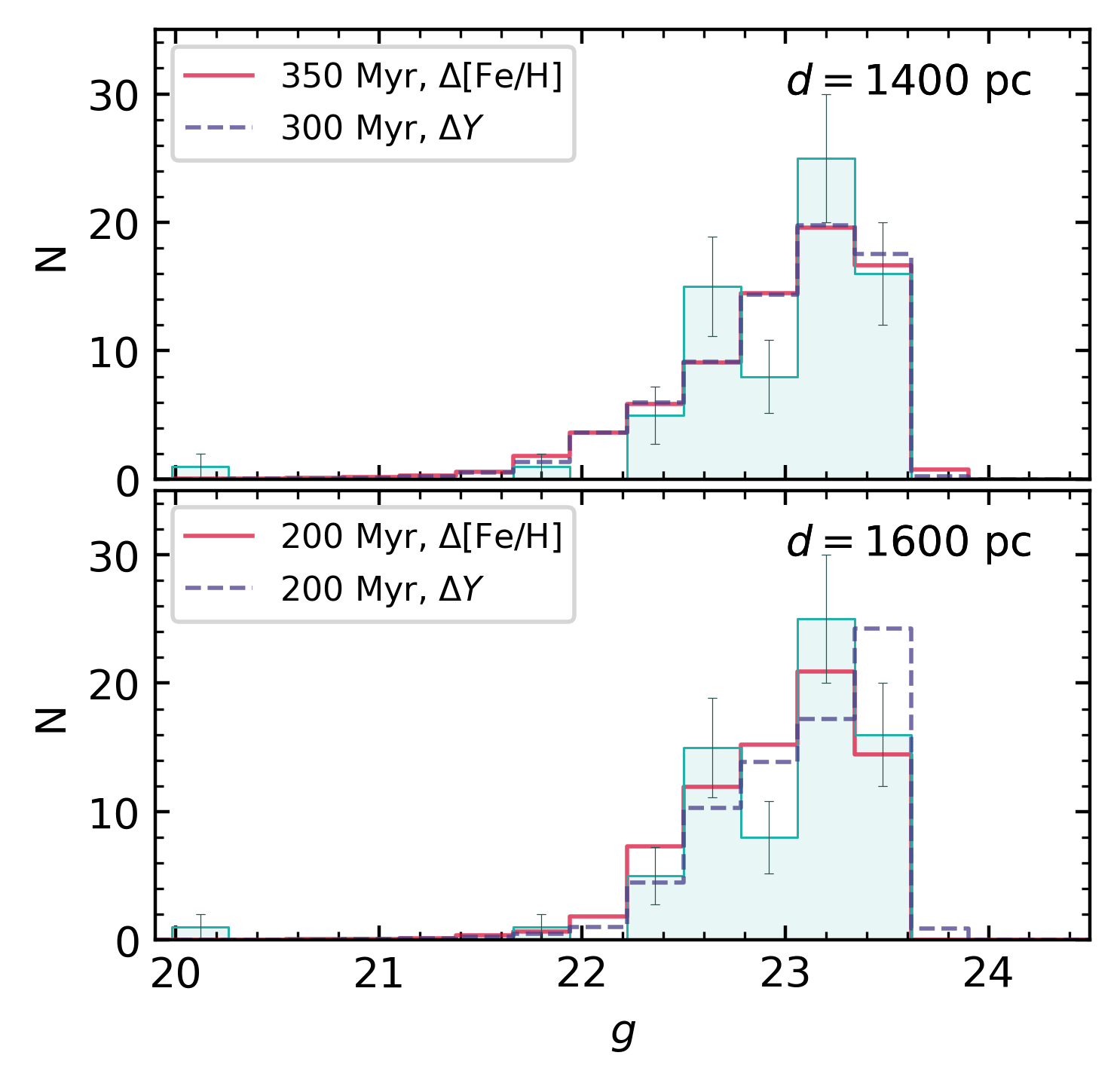}
    \caption{Completeness-corrected differential WD LF of the WDs in M37 (sea green) compared to theoretical LFs calculated 
    for the labelled ages and chemical compositions (see text for details).
    The errors in the number counts of the observed LF are also displayed.}
    \label{fig:lfth}
\end{figure}

To compare the WD $g$-band LF with models we computed grids of CO-core DA WD isochrones with the same inputs as the one in Fig.\,\ref{fig:cmd_iso} (from progenitors with $\rm [Fe/H]=0.06$), for ages between 150 and 450\,Myr at steps of 25\,Myr, and calculated 
synthetic LFs using Monte-Carlo techniques. 
As shown in Fig.\,\ref{fig:cmd_iso}, at these ages 
the WD isochrones are sequences of continuously increasing magnitude in the $g$ band, 
and the WD mass evolving at a given brightness increases monotonously with 
increasing $g$. Due to the younger ages, 
the ranges of progenitor and 
WD masses along the isochrones are narrower than in the case of globular clusters. 
At 150\,Myr the brightest part of the isochrones is populated by $\sim$\,0.95\,$M_{\odot}$ 
WDs with progenitor masses equal to $\sim$\,4.6\,$M_{\odot}$, whilst at 450\,Myr the WDs have a 
mass equal to $\sim$\,0.75\,$M_{\odot}$ with progenitors of $\sim$\,2.9\,$M_{\odot}$.
The bottom end of the isochrones is populated by 1.1\,$M_{\odot}$ WDs with $\sim$\,6.4\,$M_{\odot}$ progenitors. 

\begin{table}
    \caption{Maximum ages compatible with the WD LF cut-off magnitude, for the two distances and scenarios discussed in the text.}
    \label{ages}
    \centering
    \begin{tabularx}{.9\columnwidth}{XXX}
    \hline \hline
    $d$ (pc) & age (Myr)& scenario \\
    \hline
    1400 &  350 & $\Delta$[Fe/H] \\
    1400 &  300 & $\Delta Y$ \\
    1600 &  200 & $\Delta$[Fe/H] \\
    1600 &  200 & $\Delta Y$ \\
    \hline
    \end{tabularx}
\end{table}

For each isochrone, we have produced a sample of $g$ magnitudes of synthetic WDs (20\,000 for each age, to minimise statistical fluctuations of their magnitude distribution), 
by drawing randomly progenitor masses according to a Salpeter mass function (power law with exponent $x=-2.3$) and interpolating along the isochrone to determine the $g$ magnitude of their WD progeny. We then corrected the magnitude for the assumed cluster distance and applied a random extinction \citep[using the extinction-law by][]{zhang2023} from values of $E(B-V)$ drawn with a uniform probability within the range appropriate to 
the explored scenario ([Fe/H] or $Y$ spread).
Each synthetic $g$ was then perturbed by a random Gaussian photometric error 
with $\sigma$ estimated from the observations (see Sec.\,\ref{sec:as}).

For each of these samples (corresponding to a given WD isochrone age) we finally calculated the differential LF with the same binning of the observed one, and rescaled the total number of objects in the LF to the observed (completeness corrected) one, before comparing it with the observations. 

These sets of synthetic samples of WDs and the corresponding LFs have been computed for both the [Fe/H] spread and $Y$ spread scenarios, considering two 
distances $d$ equal to 1400 and 1600\,pc, respectively the lower 
and upper limits of the distance determination from {\it Gaia} parallaxes by \citet{gb22}.
For the assumed reference metallicity $\rm [Fe/H]=0.06$ the minimum $E(B-V)$ values (determined 
as described above) for $d=1400$\,pc are 0.26\,mag for the $\Delta$[Fe/H] scenario, and 0.31\,mag 
for the $\Delta Y$ scenario. At $d=1600$\,pc the minimum reddenings are $E(B-V)=0.31$ for the $\Delta$[Fe/H] scenario, and 0.36\,mag for the $\Delta Y$ scenario.
It is important to mention that for the $\Delta$[Fe/H] scenario we have calculated the 
WD isochrone for just one value of [Fe/H] ($\rm [Fe/H]=0.06$). This is because we have found 
that changing [Fe/H] of the progenitors by $\pm 0.20$\,dex produces isochrones virtually 
indistinguishable at these ages. The same is true also for the $\Delta Y$ scenario, with isochrones calculated considering just the minimum value of $Y$.

We have determined the oldest cluster age compatible with the 
observed WD cooling sequence, by finding the theoretical LFs that match 
the magnitude of the cut-off of the WD LF.
Fig.\,\ref{fig:lfth} shows the oldest ages compatible with the observed LF -- between 
200 and 350\,Myr, summarised in Table\,\ref{ages} -- 
for the two distances and the two scenarios discussed here. 
The derived ages are typically older (by 100-150\,Myr) for shorter distances, as expected, and at a fixed distance they are very similar in both scenarios. At these ages, all WDs along the cluster CS 
have not yet started crystallization in their CO cores.
It is important to stress that, in case the extended TO of this cluster is due to an age 
range, the WD LF tells us that the maximum age of the cluster stars cannot be older than the values given above, otherwise we should find WDs fainter than the observed LF cut-off.

\begin{figure}
    \centering
    \includegraphics[width=\columnwidth]{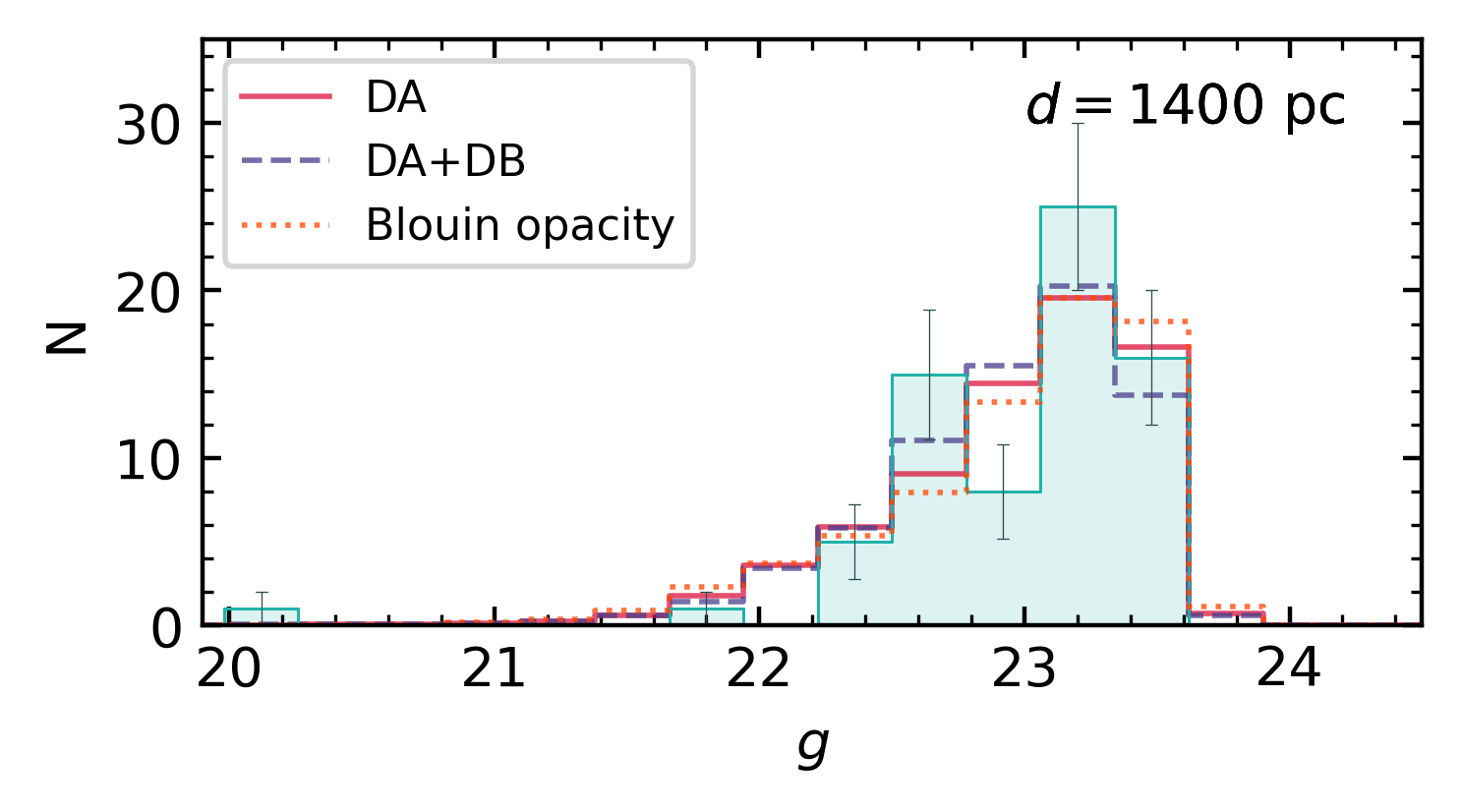}
    \caption{As the upper panel of Fig.\,\ref{fig:lfth}. The theoretical LFs are for an age 
    of 350\,Myr and correspond to the reference DA calculations of Fig.\,\ref{fig:lfth}, 
    a population of 20\% DB (helium envelope) and 80\% DA WDs, and a DA 
    population from models calculated using \citet{b20} electron conduction opacities, respectively (see text for details).}
    \label{fig:lfth_ex}
\end{figure}

We have then repeated the same procedure by employing isochrones derived from 
WD cooling models \citep[again from][]{bastiwd} calculated using the alternative \citet{b20} electron conduction opacities, and found 
results consistent with what we have previously obtained from calculations with the \citet{c07} opacities. As an example, Fig.\,\ref{fig:lfth_ex} shows how the 350\,Myr theoretical LF in the $\Delta$[Fe/H] scenario calculated using \citet{b20} opacities and a distance of 1400\,pc has the same cut-off magnitude as our reference calculations.

We have also explored the possibility that the cluster hosts not just DA WDs, but also a 20\% fraction
of WDs with He-dominated atmospheres \citep[this fraction is typical of the Galactic disc field WD
population, see, e.g.,][]{kk}. In this case, for each age, we have computed isochrones and synthetic
samples of $g$ magnitudes from the helium-envelope WD 
models by \citet{bastiwd}, and merged them with the corresponding DA samples in a proportion 20/80, before calculating the corresponding LF. The results about the 
WD-based cluster ages are again unchanged (see Fig.\,\ref{fig:lfth_ex} for an example), because in this luminosity regime H- and He-envelope WD models cool down at very similar rates. 

Finally, we have explored the role played by the adopted IFMR. For all isochrones employed in our analysis we have adopted the semiempirical \citet{ifmr} IFMR,
more specifically the one determined using the \citet{parsec} stellar evolution models \citep[see][for details]{ifmr} 
for the determination of the progenitor's lifetimes, because they are very close 
to the evolutionary lifetimes of \citet{basti} progenitors' models used 
for the calculation of the WD isochrones.
As a test, we have calculated some DA WD isochrones and LFs 
(in the $g$ band) in the age range between 
200 and 350\,Myr for $\rm [Fe/H]=0.06$, employing the \citet{ifmr} IFMR calculated using 
MIST \citep{mist} non-rotating stellar models for the progenitor lifetimes.
The effect of this alternative IFMR on the magnitude of the LF cut-off at fixed  
age is only on the order of 0.01\,mag, with a negligible impact on  
the results of our analysis.
We have repeated this same test using the independent IFMR 
determined by \citet{elbadry}, and found again a negligible impact on the 
magnitude of the theoretical LF cut-off.

\begin{figure}
    \centering
    \includegraphics[width=\columnwidth]{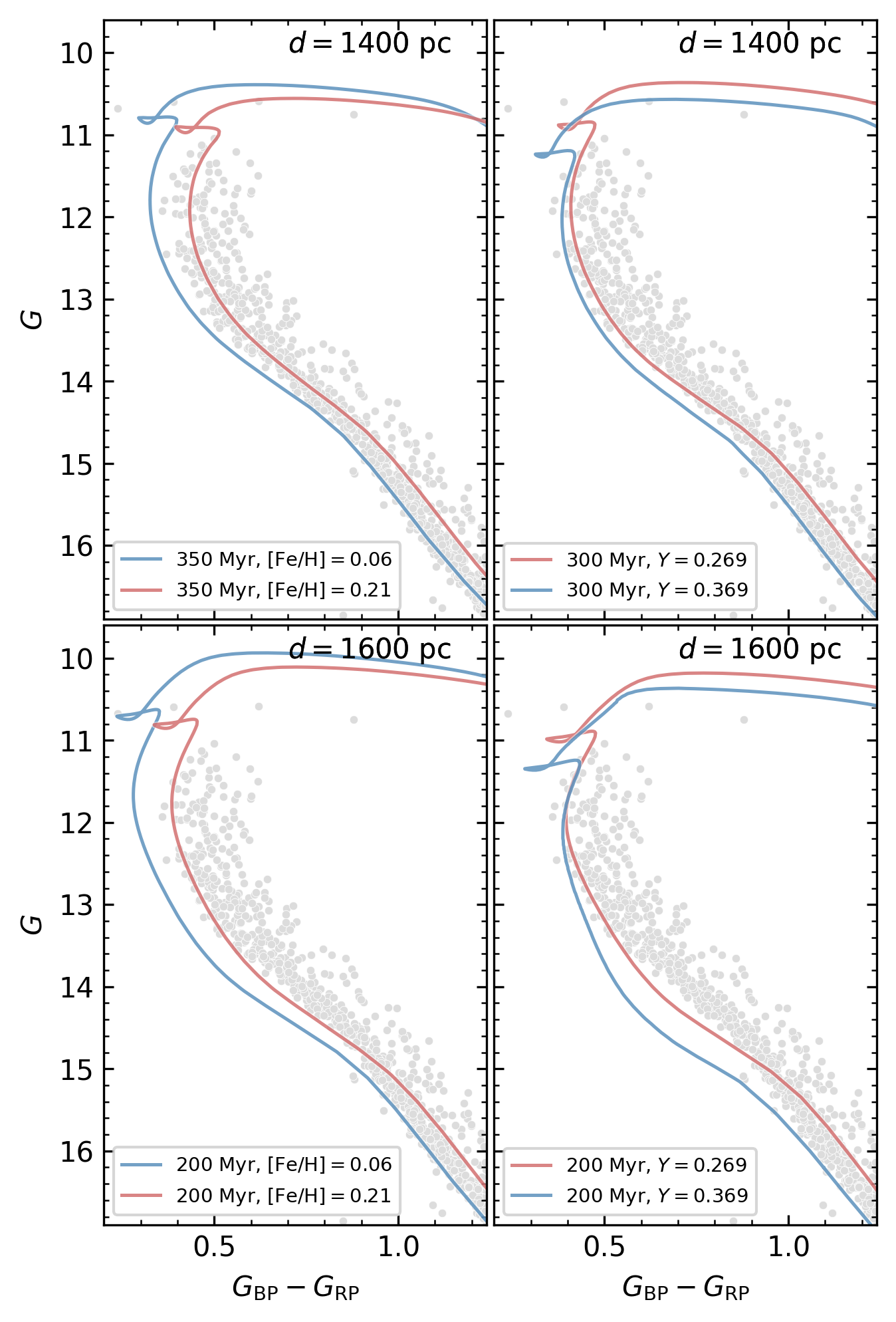}
    \caption{Cluster's {\it Gaia} CMD compared to isochrones with the labelled parameters (see text for details).}
    \label{fig:cmd_iso_gaia}
\end{figure}

\subsection{Constraints on the origin of the extended TO}

The impact of these results on the interpretation of the cluster extended TO 
is shown by 
Fig.\,\ref{fig:cmd_iso_gaia}, which is analogous to Fig.\,9 in \citet{griggiom37}. 
For each scenario and the same two distances of the WD analysis, we show here the cluster {\it Gaia} CMD 
\citep[from][]{griggiom37} together with pairs of isochrones for the combinations 
of [Fe/H] (or $Y$) and reddenings that match the blue and red
limits of the single-star sequence in the magnitude range 
studied by \citet{griggiom37}, and ages equal to the corresponding maximum ages determined from the WD LF.
According to the WD-based ages, no single star along the upper MS and TO 
can be redder than the metal richer isochrone in the $\Delta$[Fe/H] scenario, or redder than the helium poorer one in the $\Delta Y$ scenario.
This is clearly contradicted by the observed CMD, which displays large fractions 
(if not the whole cluster population) of objects redder than the reddest isochrone 
around the TO region.
This leads to the conclusion that even considering the metallicity or the helium spread derived from the unevolved MS, the ages determined from the WD LF exclude the presence of an age spread as the reason for the observed extended TO. 

\subsection{The role played by oxygen-neon core WDs}

In our analysis, we have considered the CS sequence to be populated by CO-core WDs, which are 
by far the most common type of WDs. However, according to stellar model calculations, stellar progenitors in a fairly narrow mass range  
between very approximately 6.5-7 and  9-10\,$M_{\odot}$, are expected 
to produce WDs with an oxygen-neon core and masses between $\sim$\,1.1 and 
$\sim$\,1.3\,$M_{\odot}$, originated from the electron degenerate 
cores formed at the end of core carbon burning \citep[see, e.g.,][and references therein]{siess06, poel, doh}. Predictions, both empirical and theoretical, for the IFMR of these WDs is very uncertain; however, it is still possible to make an informed assessment of their impact on the WD ages determined in our analysis. 

To this purpose, we have considered the ONe-core hydrogen-envelope WD models by \citet{camisassa} and the 
CO-core DA models from the same group \citep{camisassa17} -- both from progenitors with roughly solar metallicity -- for a strictly differential analysis using 
models calculated with the same code and physics inputs. We have considered the 1.1\,$M_{\odot}$ CO-core cooling model -- corresponding to the mass of the more massive model 
used in our WD isochrones --, and the 1.2\,$M_{\odot}$ and 1.3\,$M_{\odot}$ ONe-core models, 
and calculated WD isochrones and luminosity functions in both $\Delta$[Fe/H] and $\Delta Y$ scenarios for ages between 200 and 400\,Myr,  
using progenitors lifetimes 
from \citet{basti} and the IFMR by \citet{ifmr} for WD masses up to 
1.1\,$M_{\odot}$, as in our calculations. For the initial masses of the two ONe WD models 
we have made various assumptions, with values between 7 and 9-9.5\,$M_{\odot}$, and obtained always the same results in terms of the LF cut-off magnitudes.

We found that the ONe-core WDs are located at fainter magnitudes with respect to the 1.1\,$M_{\odot}$ CO-core objects, because of their slightly faster cooling in the relevant luminosity range; the difference (for the 1.3\,$M_{\odot}$ models) in the $g$ band LF cut-off is on the order of 0.2-0.3\,mag. This implies that including massive ONe-core WDs in the 
calculation of the isochrones would in principle reduce the age necessary to match the observed cut-off by $\sim$\,100\,Myr, thus exacerbating the inconsistency between 
WD ages and the ages required to explain the extended TO in terms of an age spread.

\section{Conclusions}
\label{conclusions}

We have presented a new {\it Sloan} photometry of the OC M37, from the 
very low-mass star regime to the main sequence TO and red clump, including 
the WD cooling sequence down to its termination.
We make publicly available these catalogue (positions, photometry, proper motions and flags) 
and the atlases, as on-line supplementary material of this article. 
We have focused our analysis on the WD CS, and determined a new, improved WD LF that we 
have exploited to set constraints on the origin of the cluster extended TO.

We have found that, irrespective of whether the chemical abundance spread revealed by 
\citet{griggiom37} photometric analysis is due to variations of [Fe/H] or $Y$, 
for the distance range determined using 
{\it Gaia} EDR3 parallaxes the ages determined from the WD LF 
are incompatible with the ages required to match the observed extended 
TO region. The maximum age allowed by the analysis of the WD LF is much 
too young compared to the age required to match the redder and fainter TO region.
This is especially true for the $Y$-spread scenario, and also for the [Fe/H]-spread 
scenario when considering the upper limit of the parallax-based distance.

Our results indirectly support the notion that stellar rotation is needed to 
explain the origin of the cluster extended TO, like the case of the OC NGC\,2818, \citet{bastian18}, where spectroscopic observations have 
confirmed the presence of a range of rotation rates among TO stars, with redder TO 
objects being faster rotators. A comprehensive analysis of the MS extended TO and 
WD cooling sequence of M37 using models including the effect of rotation\footnote{\citet{cordoni} have presented a first preliminary 
comparison of the cluster 
extended TO with models including rotation.} is now needed, together with spectroscopic measurements of the rotation velocities of TO stars, and also spectroscopic metallicities, to determine whether the abundance spread 
revealed by the photometric analysis of \citet{griggiom37} is due to a 
metal abundance or a helium spread.

\section*{Acknowledgements}
We thank our referee for comments that have helped improve the presentation of our results.
Based on observations obtained with MegaPrime/MegaCam, a joint project of CFHT and CEA/DAPNIA, at the Canada-France-Hawaii Telescope (CFHT) which is operated by the National Research Council (NRC) of Canada, the Institut National des Science de l'Univers of the Centre National de la Recherche Scientifique (CNRS) of France, and the University of Hawaii. The observations at the Canada-France-Hawaii Telescope were performed with care and respect from the summit of Maunakea which is a significant cultural and historic site.

This work has made use of data from the European Space Agency (ESA) mission
{\it Gaia} (\url{https://www.cosmos.esa.int/gaia}), processed by the {\it Gaia}
Data Processing and Analysis Consortium (DPAC,
\url{https://www.cosmos.esa.int/web/gaia/dpac/consortium}). Funding for the DPAC
has been provided by national institutions, in particular the institutions
participating in the {\it Gaia} Multilateral Agreement.

This work has also made use of observations collected at Schmidt telescopes (Asiago, Italy) of INAF.

MG, DN and LRB acknowledge support by MIUR under PRIN program \#2017Z2HSMF and by PRIN-INAF 2019 under program \#10-Bedin. 
MS acknowledges support from The Science and Technology Facilities Council Consolidated Grant ST/V00087X/1.

\section*{Data Availability}
The catalogue is available as electronic material with this paper.
The image stacks are available at \url{https://web.oapd.inaf.it/bedin/files/PAPERs_eMATERIALs/CFHT/M37_WDCS/}.

The isochrones for the MS and TO, and the WD models are available at the BaSTI-IAC model repository 
\url{http://basti-iac.oa-abruzzo.inaf.it/}. The WD models by \citet{camisassa17} and 
\citet{camisassa} are available at the La Plata group model repository \url{ http://evolgroup.fcaglp.unlp.edu.ar/TRACKS/tracks.html}.


\bibliographystyle{mnras}
\bibliography{bibliography}





\bsp	
\label{lastpage}
\end{document}